# Assessing and mitigating systematic errors in forest attribute maps utilizing harvester and airborne laser scanning data

Janne Räty*[1,2], Marius Hauglin[1], Rasmus Astrup[1], Johannes Breidenbach*[1]

[1]Norwegian Institute of Bioeconomy Research (NIBIO), Høgskoleveien 8, 1433 Ås, Norway

[2]Natural Resources Institute Finland (LUKE), Yliopistokatu 6, 80100 Joensuu, Finland

*Corresponding authors:

janne.raty@luke.fi

johannes.breidenbach@nibio.no

## Abstract

Cut-to-length harvesters collect useful information for modeling relationships between forest attributes and airborne laser scanning (ALS) data. However, harvesters operate in mature forests, which may introduce selection biases that can result in systematic errors in harvester data-based forest attribute maps. We fitted regression models (harvester models) for volume (V), height (HL), stem frequency (N), above-ground biomass, basal area, and quadratic mean diameter (QMD) using harvester and ALS data. Performances of the harvester models were evaluated using national forest inventory plots in an 8.7 Mha study area. We estimated biases of large-area synthetic estimators and compared efficiencies of model-assisted (MA) estimators with field data-based direct estimators. The harvester models performed better in productive than unproductive forests, but systematic errors occurred in both. The use of MA estimators resulted in efficiency gains that were largest for HL (relative efficiency, RE=6.0) and smallest for QMD (RE=1.5). The bias of the synthetic estimator was largest for N (39%) and smallest for V (1%). The latter was due to an overestimation of deciduous and an underestimation of spruce forests that by chance balanced. We conclude that a probability sample of reference observations may be required to ensure the unbiasedness of estimators utilizing harvester data.

# 1 Introduction

The increased availability of wall-to-wall ALS data during the past 20 years has revolutionized the mapping of forest resources (Maltamo et al. 2021). Many countries have thus established ALS-based approaches for the wall-to-wall mapping of forest resources (Asner et al. 2012; White et al. 2013; Maltamo and Packalen 2014; Nord-Larsen et al. 2017; Nilsson et al. 2017; Hauglin et al. 2021). Operationally utilized forest attribute maps are primarily constructed using the area-based approach (ABA), which relies on a statistical model using a sample of field plots and ALS data collected from an area of interest (Næsset 1997, 2014). Field data collection account for a considerable proportion of forest inventory costs, and several alternatives to reduce the inventory costs have been investigated (Noordermeer et al. 2019; Tompalski et al. 2019; Rahlf et al. 2021).

Due to high efficiency and safety for forest workers, 100 % of all commercial forest harvests in the Nordic countries are today conducted using cut-to-length harvesters. Harvesters collect huge amounts of data about the harvested trees which, so far, are under-utilized in forest inventories and the forest value chain (Kemmerer and Labelle 2020). Harvester data are a potential alternative to traditional field data in forest inventories relying on airborne laser scanning (ALS). In addition to the economic efficiency of data collection, harvester data are also attractive since these data include bucking information (Bollandsås et al. 2013; Karjalainen et al. 2019; Räty et al. 2021b). Harvesters also enable the collection of data from large area units that are too expensive to measure using traditional field measurements. For example, complete descriptions of clear-cut forest stands are valuable for the validation of remote sensing-based forest attribute maps (Vähä-Konka et al. 2020).

Several studies have suggested that harvester data are useful in ALS-based forest inventories. Hauglin et al. (2018) predicted timber volume using different regression techniques using harvester and ALS data in Norway. They compared the harvester data-based models with the models based on traditional field measurements. They reported that the root-mean-square errors (RMSEs) associated with the cross-validated timber volume predictions were similar regardless of training data. Söderberg et al. (2021) studied the operational applicability of harvester data in ALS-based forest inventories in Sweden. They predicted several forest attributes such as merchantable volume, basal area and tree size distributions using the k-nearest neighbor approach. Similarly, Maltamo et al. (2019) used harvester and ALS data in Finland to predict several forest attributes, such as diameter distributions, merchantable volume and basal area, using the k-nearest neighbor approach. Noordermeer et al. (2022) mapped and estimated merchantable timber volume at the level of grid cells and harvested sites (i.e., a cluster of adjacent stands) using harvester and ALS data. They studied the effect of tree positioning errors and the size of modeling units on the predictive performance of the models. Saukkola et al. (2019) also studied the effect of tree positioning errors on the predictive performances of the models and found that the negative effects of inaccurate tree positioning on predictive performance can be mitigated by increasing the size of modeling unit. This conclusion is in line with the findings by Noordermeer et al. (2022).

There are fundamental differences associated with the collection of harvester data and traditional field measurements, which may limit the use of harvester data in forest inventories. While traditional field plots can be spread subjectively or via a probability sample into different forest structures, harvesters are most often operating in commercially mature forests. Harvesters carry out either final fellings (i.e. clear-cuts) or a variety of thinnings. Thinnings are not common in Norway, which cause that harvesters data are typically collected from clear-cuts. Consequently, data collected by harvesters are typically biased towards commercially mature forests. Commercial maturity implies that harvested forests are harvest-ready and actively managed according to current silvicultural recommendations. Unmanaged and extensively managed forests will likely have a different forest structure that may not be well represented in the harvester data. Another limitation of harvester data is that even final harvests do not always mean that trees are completely removed from the forest stands. For example, retention trees are usually left out of the final harvests, because they can mitigate negative effects of final harvests on biodiversity (Fedrowitz et al. 2014). Retention trees and selective harvests cause incompatibilities between harvester and ALS data. In addition, small trees, that do not fulfill the size requirements of commercial timber assortments, are missing from the harvester data. Currently, harvesters do not routinely record trees left out of harvest operations. In this study, we will therefore focus on the effects of the selection bias associated with harvester data on the large-area mapping of forest attributes.

Wall-to-wall forest attribute maps are often utilized for purposes of stand-level forest management, whereas large-area estimates relying on national forest inventories (NFIs) are based on a probability sample ensuring unbiased forest attribute estimates (Kangas and Maltamo, 2006). Model-assisted (MA) estimation combines forest attribute maps and a probability sample in order to yield unbiased estimates with improved precision compared with direct estimation that exclusively utilizes field data (Saarela et al. 2015; McRoberts et al. 2017; Breidenbach et al. 2020b). If a sufficient probability sample is not available, analysts sometimes rely solely on the underlying model and summarize the pixels of a wall-to-wall forest attribute map to obtain large-area estimates (e.g. Ceccherini et al., 2020). These *synthetic estimates*, also known as "pixel-counting" estimates (Waldner and Defourny 2017), may however hold systematic errors, if the model underlying the map is insufficient (Breidenbach et al. 2021a; Palahí et al. 2021). Systematic errors in estimates hamper decision making aiming at the optimal planning and management of forest resources (Ruotsalainen et al. 2021).

Harvester data can be a cost-efficient data source for the mapping and estimation of forest attributes in ABA forest inventories relying on ALS data. However, it is not well understood if systematic errors are relevant when such forest attribute maps are used for large-area mapping and estimation. Therefore, our objectives were

(1) to study the predictive performance of models fit using harvester and ALS data (harvester models) using independent data (NFI field plots),

(2) to study the magnitude of systematic errors in synthetic estimates that would result from forest attribute maps based on harvester models,

(3) to compare the precisions of MA estimates (correction of synthetic estimates using NFI data) to direct estimates (only NFI data) in an 8.7 Mha study area in Norway.

The forest attributes of interest in this study were timber volume (V), Lorey's height (HL), stem frequency (N), above-ground biomass (AGB), basal area (G), and quadratic mean diameter (QMD).

## 2 Material and Methods

### 2.1 Study area

The 8.7 Mha study area is located between the latitudes of 58.0°N and 66.1°N in Norway (**Figure** ) and comprises forests below the coniferous limit defined as the lowland stratum in the Norwegian NFI (Breidenbach et al. 2020a). Forests cover 84 % of the study area according to the NFI definition of the ability to reach 10 % crown cover and 5 m height. Norway spruce (*Picea abies*

[L.] Karst.) is the dominating tree species and comprise 48 % of total volume in the study area. Scots pine (*Pinus sylvestris* [L.]) is especially common on sites that are not fertile enough for Norway spruce. Deciduous trees, mostly birch (*Betula spp.*) typically grow as mixtures among coniferous species. Scots pine and deciduous species comprise 33 % and 19 % of total volume, respectively.

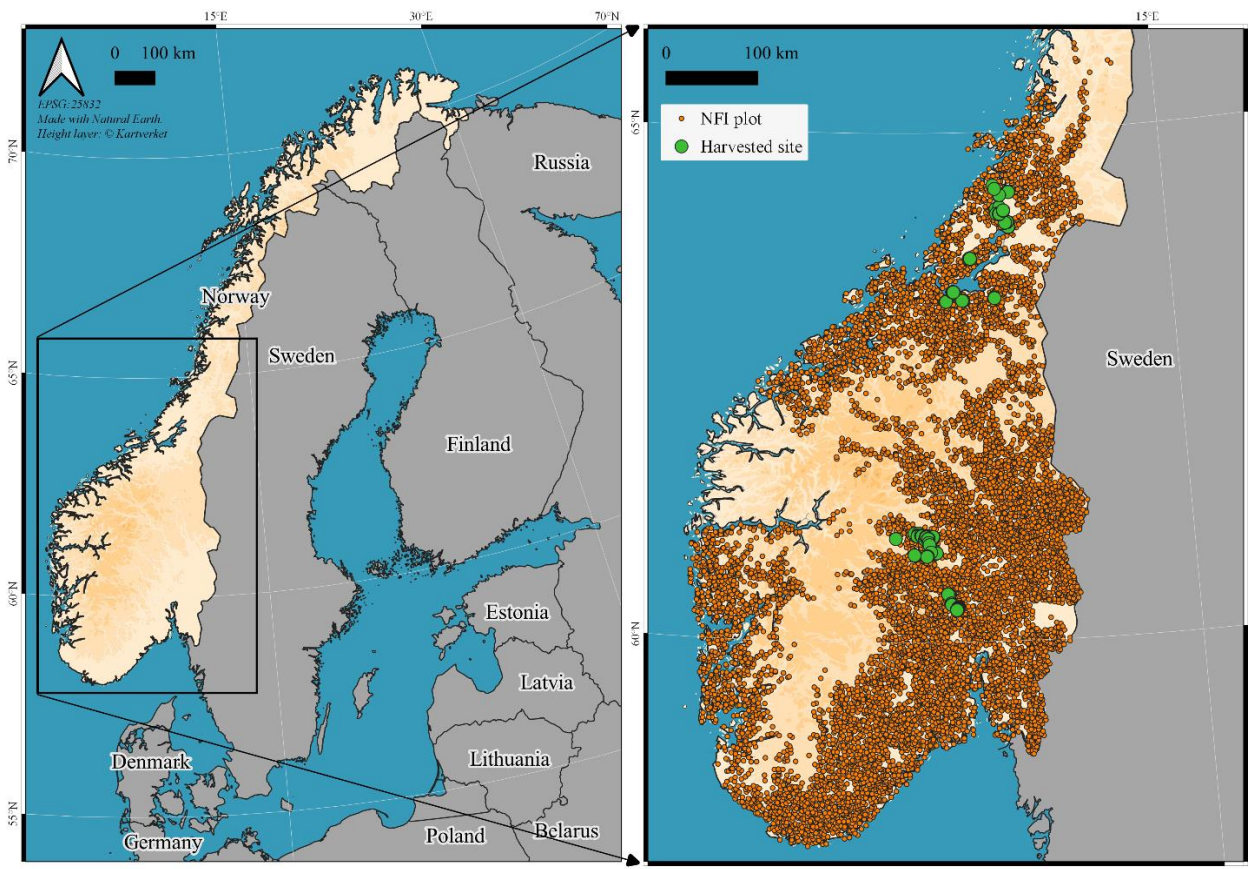

**Figure 1.** Approximate locations of national forest inventory (NFI) plots and harvested sites in the study area. Country vectors from Natural Earth (2021) and the basemap of Norway from Kartverket (2021). Map was created using the QGIS software (QGIS Development Team 2022).

## 2.2 National forest inventory data

The Norwegian NFI follows a systematic grid of 3 km × 3 km in the lowland stratum and the plots are located at the grid knots. The NFI plots are circular with a fixed area of 250 m$^2$, where all trees with diameter at breast height (DBH) ≥ 5 cm are recorded. Tree heights are recorded for a subset of approximately 10 trees per plot. The heights of remaining trees, as well as volume and biomass for each tree, are predicted using tree-level height-diameter, V, and AGB models. A detailed description of the sampling design and measurement protocol of the Norwegian NFI is given by (Breidenbach et al. 2020a).

In this study, the plot-level forest attributes HL, V, N, AGB, G, and QMD were variables of interest. The stand-level attributes site index, i.e. height classes in the age of 40 (Tveite 1977; Tveite and Braastad 1981) and the Norwegian NFI's maturity classification (M1, …, M5) (Breidenbach et al. 2020a) were used to specify data domains in which model performances were evaluated. The maturity classification indicates the development stage of forest which depends on age, site index and tree species. This means that highly productive forests will achieve the status of an older maturity at younger age than low-productive forest. The maturity classes can be categorized into overlapping age groups as follows:

- M1: Forest under regeneration (age 0 yrs.)
- M2: Young forest (age 1–54 yrs.)
- M3: Young production forest (age 15–84 yrs.)
- M4: Older production forest (age 25–119 yrs.)
- M5: Mature forest (minimum age 40–120 yrs.)

The age groups for a given site index class do not overlap. Typically, most of the commercial harvests in Norway are expected to occur in maturity classes M4 and M5.

The NFI data within the study area comprise of 9,615 plots that were measured between 2016 and 2020. We excluded 305 plots where harvests occurred between field measurements and ALS data acquisition. The 9,310 plots covering all land uses constitute estimation data that were used in large-area estimation (Section 2.7). A total of 7,841 plots were located in forest.

The NFI data were used to establish three datasets to evaluate the predictive performance of the harvester models: (1) a dataset that included all forest plots (ALL), (2) a dataset that included productive forests (PROD), and a dataset that included unproductive forests (UPROD). Forests under regeneration (M1) were not included in the datasets because they are typically covered by trees below the caliper threshold. We utilized attributes assessed in the NFI to decide which dataset a plot belongs to. Site index (> 11), coniferous volume proportion (> 80 %) and forwarding distance to the closest road (< 500 m) were used as criteria to determine the PROD dataset. These are plots in productive and managed forests which are likely to be clear felled by harvesters in the future. Site index (< 11), coniferous volume proportion (< 60 %) and forwarding distance to the closest road (> 500 m) were used to determine the UPROD dataset. The datasets only included plots that were completely on forest land. The ALL, PROD, and UPROD datasets comprise 5,858, 614, and 274 NFI plots, respectively. Statistics associated with the datasets are shown in Table 1. We present the statistics by maturity classes since the maturity classification was utilized in the performance evaluation of the models.

**Table 1.** Means and standard deviations (SDs) of forest attributes in the test data (ALL, PROD and UPROD) that consist of national forest inventory (NFI) data. The ALL, PROD and UPROD datasets comprise all forest, managed forest plots, and unmanaged forest plots, respectively. V –

volume, HL – Lorey's height, N – stem frequency, AGB – above-ground biomass, G – basal area, QMD – quadratic mean diameter, M2, …, M5 – NFI maturity classification (see section 2.2).

| Data | | ALL | PROD | UPROD | ALL | PROD | UPROD | ALL | PROD | UPROD | ALL | PROD | UPROD | ALL | PROD | UPROD |
|---|---|---|---|---|---|---|---|---|---|---|---|---|---|---|---|---|
| Maturity class | | | M2–M5 | | | M2 | | | M3 | | | M4 | | | M5 | |
| HL (m) | Mean | 13.4 | 15.8 | 9.5 | 9.6 | 9.2 | 9.2 | 12.3 | 13.9 | 7.8 | 14.9 | 18.8 | 9.3 | 14.8 | 22.1 | 10 |
| | SD | 4.4 | 4.8 | 2.6 | 3.7 | 3.6 | 2.8 | 3.2 | 3.0 | 2.5 | 4.3 | 3.0 | 2.3 | 4.2 | 4.0 | 2.5 |
| V (m³ · ha⁻¹) | Mean | 149.4 | 226.2 | 62.7 | 41.7 | 49.1 | 26.3 | 134.6 | 175.6 | 42.7 | 199.9 | 309 | 68.4 | 172.2 | 398.3 | 76.5 |
| | SD | 123 | 154.9 | 44.2 | 40.2 | 48.2 | 24.2 | 88.9 | 97.5 | 45.7 | 143.2 | 140.9 | 37.3 | 123.9 | 188 | 43.1 |
| N (stems · ha⁻¹) | Mean | 1,082 | 1,280 | 1,014 | 734 | 906 | 452.5 | 1,473 | 1,478 | 1,121 | 1,298 | 1,222 | 1,465 | 885 | 998 | 1,032 |
| | SD | 696 | 638 | 624.4 | 595 | 525 | 426.6 | 779 | 670 | 616.4 | 739 | 562 | 807.2 | 492 | 540 | 464.9 |
| AGB (t · ha⁻¹) | Mean | 20.8 | 27.0 | 46.2 | 7.5 | 8.9 | 19.4 | 20.5 | 24.5 | 29.9 | 25.8 | 33.7 | 49.5 | 23.2 | 38 | 56.9 |
| | SD | 12.6 | 13.9 | 32.5 | 5.9 | 6.6 | 18.6 | 10.3 | 10.4 | 31 | 13.5 | 12.4 | 26.8 | 11.8 | 15.4 | 32.1 |
| G (m2 · ha⁻¹) | Mean | 95.7 | 138.4 | 12.5 | 28.2 | 33.3 | 5.4 | 87.1 | 112.2 | 9 | 125.1 | 183.9 | 14.1 | 110.8 | 235.7 | 14.9 |
| | SD | 74.1 | 90.4 | 7.4 | 25.9 | 30.1 | 4.5 | 55.3 | 61.2 | 7.3 | 84.2 | 81.9 | 7.1 | 74 | 107.8 | 6.6 |
| QMD (cm) | Mean | 16.3 | 16.8 | 13.1 | 11.9 | 10.9 | 14.1 | 13.8 | 15.0 | 9.7 | 16.6 | 19.6 | 11.4 | 19.1 | 23.3 | 14 |
| | SD | 5.6 | 5.3 | 4 | 4.7 | 4.0 | 5.8 | 3.7 | 3.7 | 2.2 | 4.6 | 4.1 | 2.6 | 5.6 | 5.6 | 3.5 |
| No. of plots | n | 5,858 | 614 | 274 | 888 | 73 | 48 | 1,353 | 271 | 29 | 1,197 | 222 | 47 | 2,420 | 48 | 150 |

## 2.3 Harvester data

Harvester data were collected using four different machines between 2019 and 2021. The harvester data were collected from harvested sites that are different-sized harvest units determined by the harvesting companies. The harvested sites were clusters of adjacent final-felled stands (i.e., clear-cuts with possible retention trees) and located between latitudes 60.3 N and 64.5 N (Figure 1). Harvested sites were typically dominated by Norway spruce or Scots pine. Each harvester was equipped with a GNSS (global navigation satellite system) receiver that recorded a geographical location when a tree was felled. Three of the harvesters were capable of determining the position of the boom tip with sensors recording crane length and orientation, whereas one harvester only registered machine positions. The geographical locations of trees that were based on machine positions were stripe-patterned on the site and were post-processed by distributing the XY positions by adding a random value of 8 m to both X and Y coordinates (Räty et al. 2021b). The positioning errors associated with the harvested trees are assumed to vary between 5–20 m.

The harvested sites were overlaid with the stand-like segments of the forest resource map SR16 (Astrup et al. 2019; Hauglin et al. 2021). Since the harvested sites did not always follow the SR16 segmentation, we first delineated the harvested sites using an alpha shape approach ($\alpha = 25$ m) that created segments based on the XY positions of the harvested trees. Finally, each SR16 segment was cropped with the corresponding alpha shape to establish a *harvested segment*. More information on the generation of harvested segments can be found in Räty et al. (2021b). In total, 277 harvested segments were further split into 1,024 m$^2$ units (*harvested grid cells*) that were used as modeling units. The grid cell size was selected based on prior knowledge on the effect of modeling unit size when using harvester data. Saukkola et al. (2019) suggested that expanding the grid cell size mitigates the negative effects inherited from the uncertain XY positioning of harvested trees. We accepted all grid cells that had at minimum 80 % of area inside the boundaries of harvested segments. A total of 1,608 harvested grid cells containing 118,699 harvested trees were utilized in this study. 84, 14 and 2 % of the harvested grid cells were dominated by Norway spruce, Scots pine, and deciduous species, respectively.

The harvester data were saved in the Standard for Forest machine Data and communication (StandForD2010) format (Arlinger et al. 2012). The harvesters registered diameter measurements for every 10 cm position along stem, log product lengths and log product volumes (Nordström and Hemmingsson 2018), but they did not register the timber volume associated with treetops. The following steps were taken to overcome this limitation (Hauglin et al. 2018; Räty et al. 2021b): tree height was predicted using species-specific height-diameter curves (Eide 1954; Strand 1967; Blingsmo 1985) that were iteratively calibrated according to harvester-based diameter measurements along the stem. The harvester-based diameter measurements were also used to predict the DBH for each tree. The harvester-based DBH and tree height were used as input parameters in species-specific volume functions (Breidenbach et al. 2020a) to obtain total volume for each harvested tree.

AGB was predicted for each tree with species-specific biomass functions used in the Norwegian NFI (Breidenbach et al. 2020a). Harvester-based DBH and tree height were used as input parameters in the models.

The forest attributes V, HL, N, AGB, G, and QMD were calculated at the level of harvested grid cells. Predicted tree heights were used for HL, whereas G and QMD were calculated based on harvester-based DBHs. Table 2 shows means and standard deviations associated with the forest attributes in the harvested grid cells. Figure 2 shows the DBH distribution of all harvested trees in the harvested grid cells.

**Table 2.** Means and standard deviations (SDs) of forest attributes in the harvested grid cells (n = 1,608).

|  | Mean | Standard deviation |
|---|---|---|
| HL (m) | 16.8 | 2.9 |
| V (m$^3 \cdot$ ha$^{-1}$) | 210.4 | 114.9 |
| N (stems $\cdot$ ha$^{-1}$) | 760 | 398 |
| AGB (t · ha⁻¹) | 26.0 | 11.3 |
| G (m2 · ha⁻¹) | 134.5 | 67.6 |
| QMD (cm) | 21.5 | 3.9 |

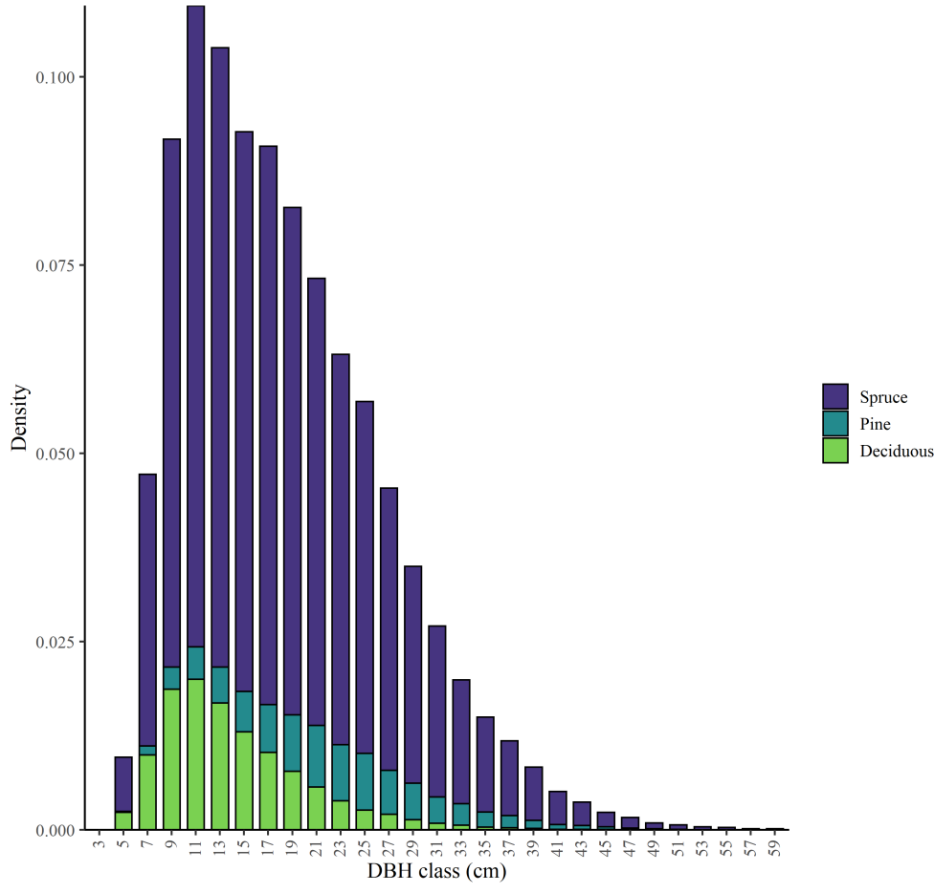

**Figure 2.** Diameter at breast height (DBH) distribution of trees in the harvester data.

## 2.4 Airborne laser scanning data

ALS data were collected in several flight campaigns between 2010 and 2018 (Hauglin et al. 2021). The flight parameters differed among the ALS campaigns and the resulting mean point densities varied between 2–5 points per square meter. A digital terrain model (DTM, 1 m × 1 m) was created using the returns classified as ground hits. Processing and classification of the ALS point clouds were carried out by the contractor for each ALS campaign, following specifications from the Norwegian Mapping Authority (Kartverket 2019). The DTM was subtracted from the orthometric elevation measurements of the ALS data to obtain normalized height measurements. The normalized ALS data were overlaid with the NFI plots and harvested grid cells in order to calculate height mean, variance, and percentiles (10th, 25th, 50th, 75th, and 95th). We also calculated a density metric (d2), which indicates the proportion of echoes above 2 m. We only calculated metrics based on the first echoes. The calculated ALS metrics are given in Table 3.

**Table 3**. Metrics calculated from the height measurements of airborne laser scanning (ALS) data.

| Metric | Description |
| --- | --- |
| hmean | Mean of ALS height measurements. |
| hvar | Variance of ALS height measurements. |
| h10, ..., h95 | Percentile calculated from ALS height measurements. Percentiles: 10th, 25th, 50th, 70th and 95th. |
| d2 | Proportion of ALS height measurements above 2 m threshold |

## 2.5 Study workflow

This study consisted of four steps (Figure 3). (1) Harvester models were fitted using harvester and ALS data. (2) The models were used to predict forest attributes at the NFI data. (3) The predictions were evaluated per maturity class and dominant tree species in all forest plots (ALL) and per maturity class in productive and unproductive forests (PROD and UPROD). (4) Evaluations of large-area MA estimation and systematic errors of synthetic estimates were carried out using the NFI plots in the 8.7 Mha study area. A full-scale wall-to-wall mapping using harvester models was not needed for this exercise.

Each step of the workflow is explained in detail in Sections 2.6 and 2.7.

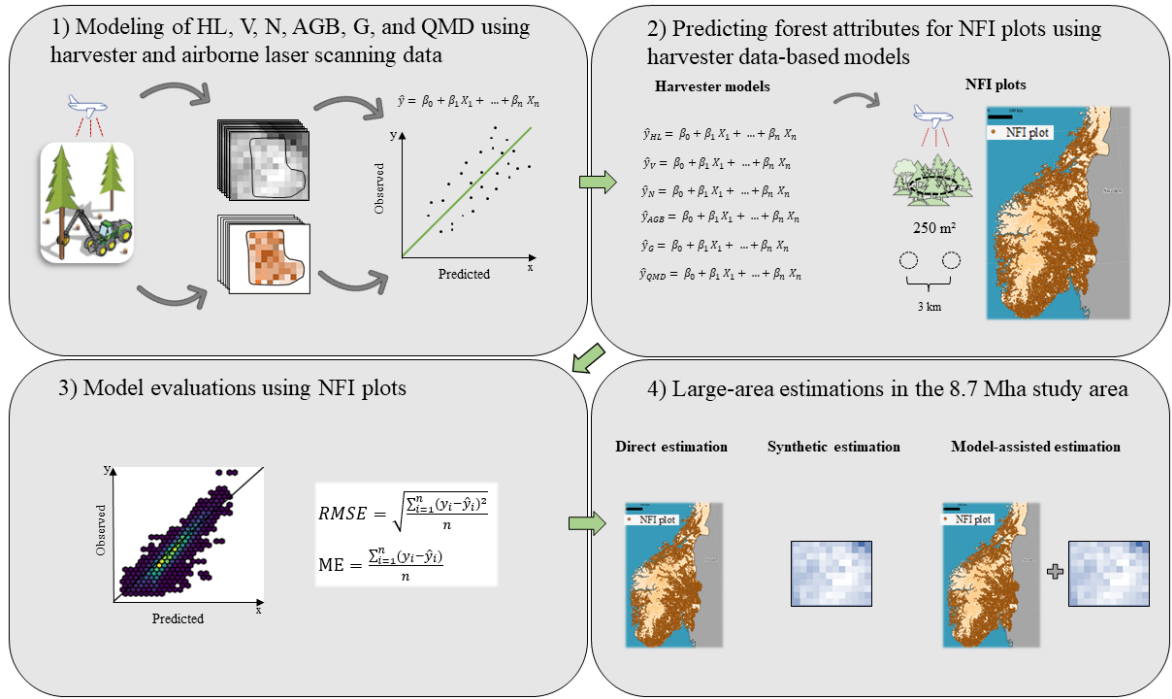

**Figure 3.** Study workflow. NFI – National forest inventory, HL – Lorey's height, V – volume, N – stem frequency, AGB –above-ground biomass, G – basal area, QMD – quadratic mean diameter, RMSE – root-mean-squared error, ME – mean error.

## 2.6 Modeling of forest attributes

### 2.6.1 Harvester models

Coefficients of linear regression models were estimated using ordinary least squares (OLS) technique. The OLS models have performed well compared with well-known machine learning approaches, such as k-nearest neighbor and Random Forests (Cosenza et al. 2021). In this study, the OLS model's capability to extrapolate beyond the training data is a desired characteristic in the model application phase. The training data consisted of the harvested grid cells with harvester-collected tree measurements and ALS metrics.

We manually selected the predictor variables based on subject-matter theory and our prior knowledge on the forest attribute modeling in the study area. We also took into account the resolution mismatch between the training and test data, which is a possible source of systematic errors in the large-area applications that aim to yield unbiased estimates (Packalen et al. 2019). In order to minimize the scaling effect, we preferred predictor variables that have resolution invariant means (e.g. height mean and proportions of echoes above a fixed threshold) and did not use nonlinear transformation on response variables. It must be noted that the theory of resolution invariant means holds in conditions with constant number of echoes per plot and when the forest attribute is additive in nature. Some forest attributes, namely LH and QMD in our study, are only additive under the constraint of a constant number of trees per area unit (Packalen et al. 2019).

The time lag between the collection of ALS data and harvester data was on average of 6 growing seasons and ranged between 2 and 9 growing seasons. This could cause systematic errors when harvester models are applied to independent data. The time lag between the acquisition of ALS data and field measurements was therefore incorporated as a predictor variable ($time_{diff}$) in the models as also described by (Hauglin et al. 2021). The $time_{diff}$ predictor variable accounts for the mismatch between the ALS and field data acquisition and denotes the number of growing seasons between the ALS data acquisition and field measurements. The expected effect of the $time_{diff}$ variable in the forest attribute models, except the model for N, is to obtain a positive coefficient estimate which indicates that the predicted values increase when $time_{diff}$ increases. We did not employ the $time_{diff}$ variable if the interpretation of the coefficient estimate was not reasonable according to subject-matter knowledge, i.e. in case where the coefficient associated with the $time_{diff}$ variable was negative.

### 2.6.2 Performance assessments

Predictive performances of the forest attribute models were evaluated using predicted-$R^2$ (Eq. 1), root-mean-square error (RMSE, Eq. 2), and mean error (ME, Eq. 3).

$$\text{predicted-}R^2 = 1 - \frac{\sum_{i=1}^{n}(y_i-\hat{y}_i)^2}{\sum_{i=1}^{n}(y_i-\bar{y})^2} \quad (1)$$

where $y_i$ and $\hat{y}_i$ are observed and predicted values in unit i. The number of units is denoted by $n$. For simplicity, we will refer to the predicted-$R^2$ as $R^2$.

$$RMSE = \sqrt{\frac{\sum_{i=1}^{n}(y_i-\hat{y}_i)^2}{n}} \quad (2)$$

$$ME = \frac{\sum_{i=1}^{n}(y_i-\hat{y}_i)}{n} \quad (3)$$

The $R^2$, RMSE and ME values were calculated both in the training data and test data. We use the subscript "TRAIN" (e.g. $RMSE_{TRAIN}$) to indicate that training data were used in the model evaluation. Whenever necessary, we also use subscripts ALL and PROD for performance measures to distinguish test data. The relative RMSE (RMSE%) and ME (ME%) were calculated by dividing absolute value by the observed mean of forest attribute. The predictive performances of the models were also visually evaluated by observed vs predicted figures in the training and test data.

## 2.7 Large-area estimation of forest attributes

### 2.7.1 Direct estimation

Direct estimation refers to the case in which forest attribute estimates and their variances are calculated based on the NFI data. Our interest was on the forested land, which means that we calculated the estimates as a ratio of forest attribute and forest land area estimates (Mandallaz, 2008, p. 63). The NFI plots that had their plot center in forest were considered as forest land plots. In case of the non-additive attribute HL, forest land refers to forest land with measured trees.

The direct estimator for mean forest attribute $\hat{\mu}$ per forest land is

$$\hat{\mu} = \frac{\sum_{j \in S} y_j I_j}{\sum_{j \in S} I_j} \qquad (4)$$

where $I_j$ is an indicator variable that takes a value 1 when NFI plot $j$ is on forest land and 0 on non-forest land. $S$ refers to the probability sample in the study area.

There exists no design-unbiased variance estimator available for systematic sampling designs. However, simulation studies and empirical analysis have shown that a variance estimator assuming simple random sampling (SRS) provides acceptable variance estimates (always conservative) also under systematic sampling designs in the case of low sampling proportions that are typical for NFIs (Magnussen et al. 2020; Räty et al. 2021a). A variance estimator for $\hat{\mu}$ assuming SRS (Cochran 1977, sec. 6.4; Mandallaz 2008, p. 63) is

$$\widehat{var}(\hat{\mu}) = \frac{1}{\left(\frac{\sum_{j \in S} I_j}{n_S}\right)^2} \frac{1}{n_S(n_S-1)} \sum_{j \in S} (z_j)^2 \qquad (5)$$

where $z_j = (y_j - \hat{\mu})I_j$ and $n_S$ refers to the number of NFI plots in the sample $S$.

The standard error (SE) of an estimate is

$$SE(\hat{\mu}) = \sqrt{\widehat{var}(\hat{\mu})} \qquad (6)$$

We used $2 \times SE$ to evaluate the precision of an estimate.

### 2.7.2 Model-assisted estimation

The MA estimator comprises of two components: (1) the synthetic estimate $\hat{\mu}_{syn}$ and (2) an estimate of a correction factor ($\hat{\mu}_{cor}$). The $\hat{\mu}_{cor}$ component is used to correct the systematic error associated with the model predictions.

The MA estimator is

$$\hat{\mu}_{MA} = \hat{\mu}_{syn} + \hat{\mu}_{cor} \qquad (7)$$

where $\hat{\mu}_{cor} = \frac{\sum_{j \in S} e_j I_j}{\sum_{j \in S} I_j}$ in which $e_j$ refers to the residual $(y_j - \hat{y}_j)$ at NFI plot $j$ (Breidenbach et al. 2021b).

The variances of the MA estimates were calculated using the SRS estimator (Eq. 4) with $z_j = (e_j - \bar{e}) I_j$ where $\bar{e}$ is the mean of model residuals observed in the NFI plots of interest. Note that the map-based estimate $\hat{\mu}_{syn}$ is not needed to evaluate variance estimates. We predicted the forest attributes for the NFI plots using the harvester models to obtain $e_j$. Observed forest land classification was used in the MA estimation, which means that we assumed an error-free classification of forest land in the study area.

### 2.7.3 Evaluating efficiencies of estimators

The efficiencies of the estimators were evaluated using SE, and relative efficiency (RE). The RE allows a comparison of the direct and MA estimators and is given by

$$RE = \frac{\widehat{var}(\hat{\mu})}{\widehat{var}(\hat{\mu}_{MA})} \qquad (8)$$

The RE values allow a straightforward comparison between the efficiencies of the estimators, and an RE value larger than 1 indicates that the MA estimator is more efficient than the direct estimator. Assuming SRS, the number of sample plots required to achieve the same variance as the MA estimate can be calculated by means of RE. Given that the MA estimation determines the desired variance, the similar variance can be achieved using the direct estimator by increasing $n_S$ (the number of sample plots) by $n_S \times RE$.

# 3 Results

## 3.1 Predictive performance of the harvester models using the training data

Figure 4 shows observed versus predicted values associated with the model predictions in the training data. The model coefficients, RMSE%$_{TRAIN}$, and $R^2$ values associated with the harvester models are given in Table 4. The RMSE%$_{TRAIN}$ values associated with the model predictions were 6.8, 30.3, and 39.0 % for HL, V, and N, respectively. Correspondingly, the $R^2$ value was largest for HL ($R^2$=0.85), followed by V ($R^2$=0.69) and N ($R^2$=0.44). The estimated model coefficients and their standard errors are in Table 4. The results associated with AGB, G, and QMD are shown in the Appendix to simplify the presentation and because they were comparable to the results for HL, V and N.

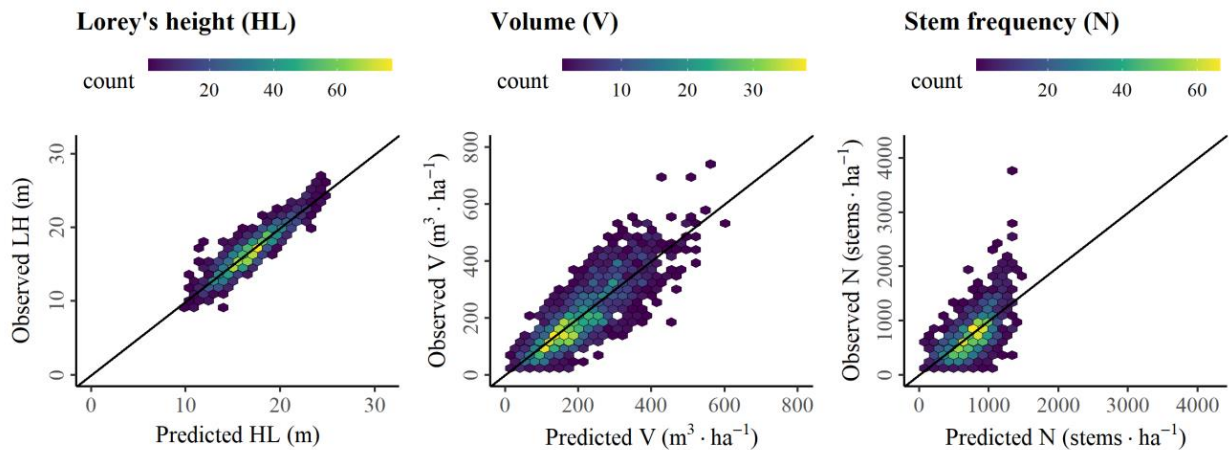

**Figure 4.** Observed versus predicted values associated with the harvester models in the training data.

**Table 4.** Estimated coefficients, standard errors (SEs) of coefficients, p-values (p), relative root-mean-square errors (RMSE%s) and coefficients of determination ($R^2$s) associated with the forest attribute models fit using harvester and airborne laser scanning data.

| | Lorey's height | Volume | Stem frequency |
|---|---|---|---|

| Predictors | Estimates | SE | p | Estimates | SE | p | Estimates | SE | p |
|---|---|---|---|---|---|---|---|---|---|
| (Intercept) | 3.77 | 0.17 | <0.001 | -76.09 | 8.28 | <0.001 | -497.48 | 36.17 | <0.001 |
| h95 | 0.77 | 0.01 | <0.001 | | | | | | |
| hmean | -0.04 | 0.02 | 0.016 | 33.84 | 1.00 | <0.001 | -48.10 | 4.46 | <0.001 |
| $time_{diff}$ | 0.19 | 0.01 | <0.001 | 9.23 | 0.67 | <0.001 | | | |
| d2 | | | | -6.14 | 18.95 | 0.746 | 2426.77 | 84.97 | <0.001 |
| $RMSE\%_{TRAIN}$ | 6.75 | | | 30.30 | | | 39.03 | | |
| $R^2$ | 0.85 | | | 0.69 | | | 0.44 | | |

## 3.2 Predictive performance of the harvester models using the test data

Figure 5 shows the observed versus predicted values for the plots of ALL data dominated by spruce, pine, and deciduous species. The harvester models performed better in coniferous-dominated than deciduous-dominated forests. The HL and V predictions associated with deciduous-dominated plots were considerably overpredicted. For all attributes (HL, V, and N), the smallest RMSE% values were observed in spruce dominated forests. The harvester model for N tended to yield underpredictions regardless of dominant tree species, while the most considerable underpredictions were observed in plots with large stem frequency, i.e. typically young forests (see also Figure 6).

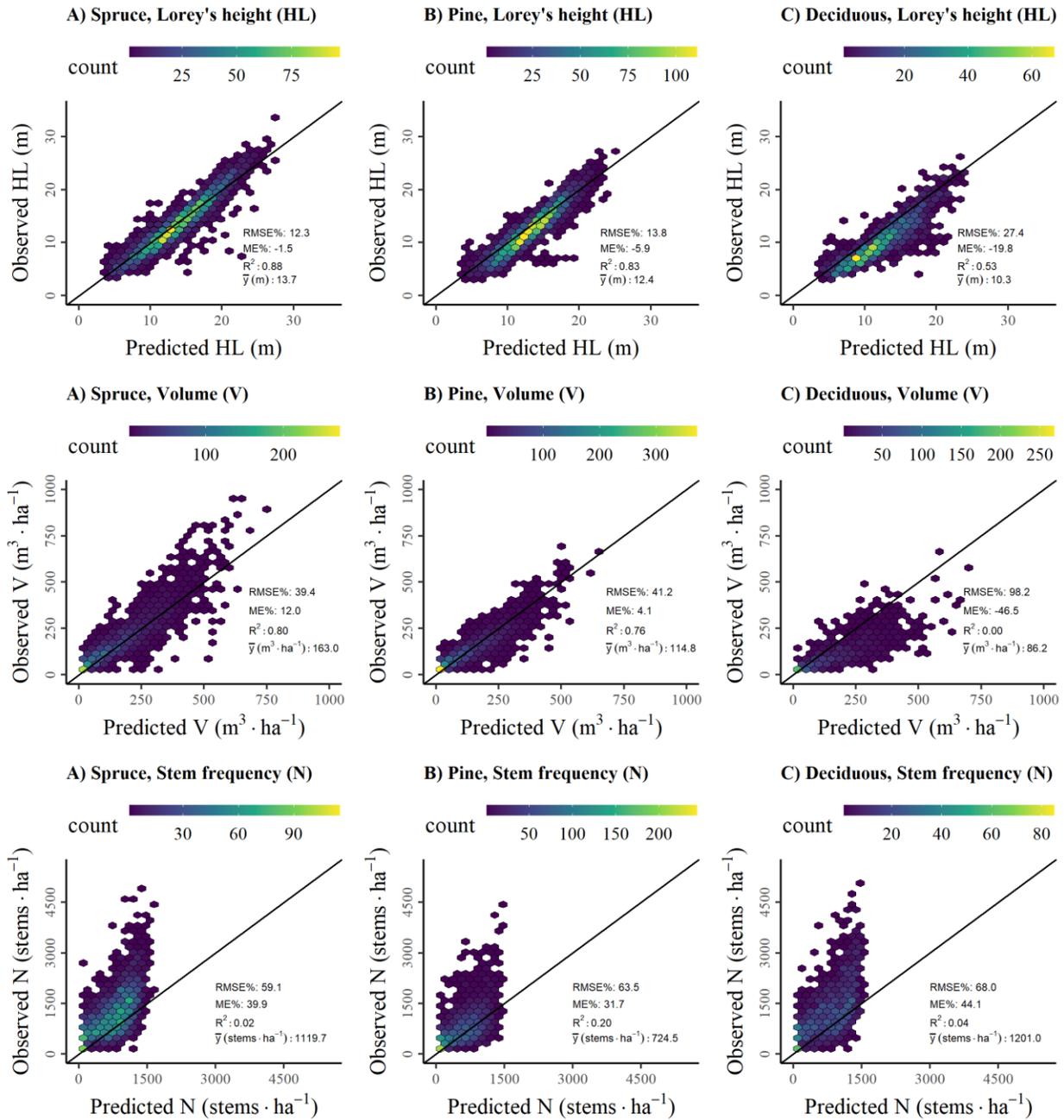

**Figure 5**. Observed versus predicted values by dominant tree species (Spruce: Norway spruce, Pine: Scots pine, Deciduous: Deciduous species) in the ALL dataset.

Figure 6 shows the RMSE% and ME% values associated with the predictions in the ALL, PROD, and UPROD datasets by maturity classes M2, …, M5 (and all maturity classes M2–M5). The RMSE%$_{PROD}$ values were always smaller than the RMSE%$_{ALL}$ values, whereas RMSE%$_{UPROD}$ values were largest in all cases. This indicates that the harvester models performed better in productive than unproductive forests. The harvester models resulted in the smallest RMSE%$_{PROD}$ values in maturity classes M4 and M5. This confirms the expectation that the larger sample of harvester data in mature productive forests results in better models for these forest types.

The ME% values associated with V were close to zero in the maturity classes M3, M4, and M5 in the dataset ALL. The ME% value was also close to zero (-0.8 %) when all maturity classes M2–M5 were considered. The ME% values associated with the N and V models were, however, considerable (|ME%| > 24 %) in maturity class M2 indicating severe systematic prediction errors. The results associated with the harvester model of N indicated that the smallest systematic errors were in old forests (M5), although the ME% values associated with N were considerable regardless of maturity class and dataset. The |ME%| values associated with the UPROD dataset were always larger than those of the PROD and ALL datasets.

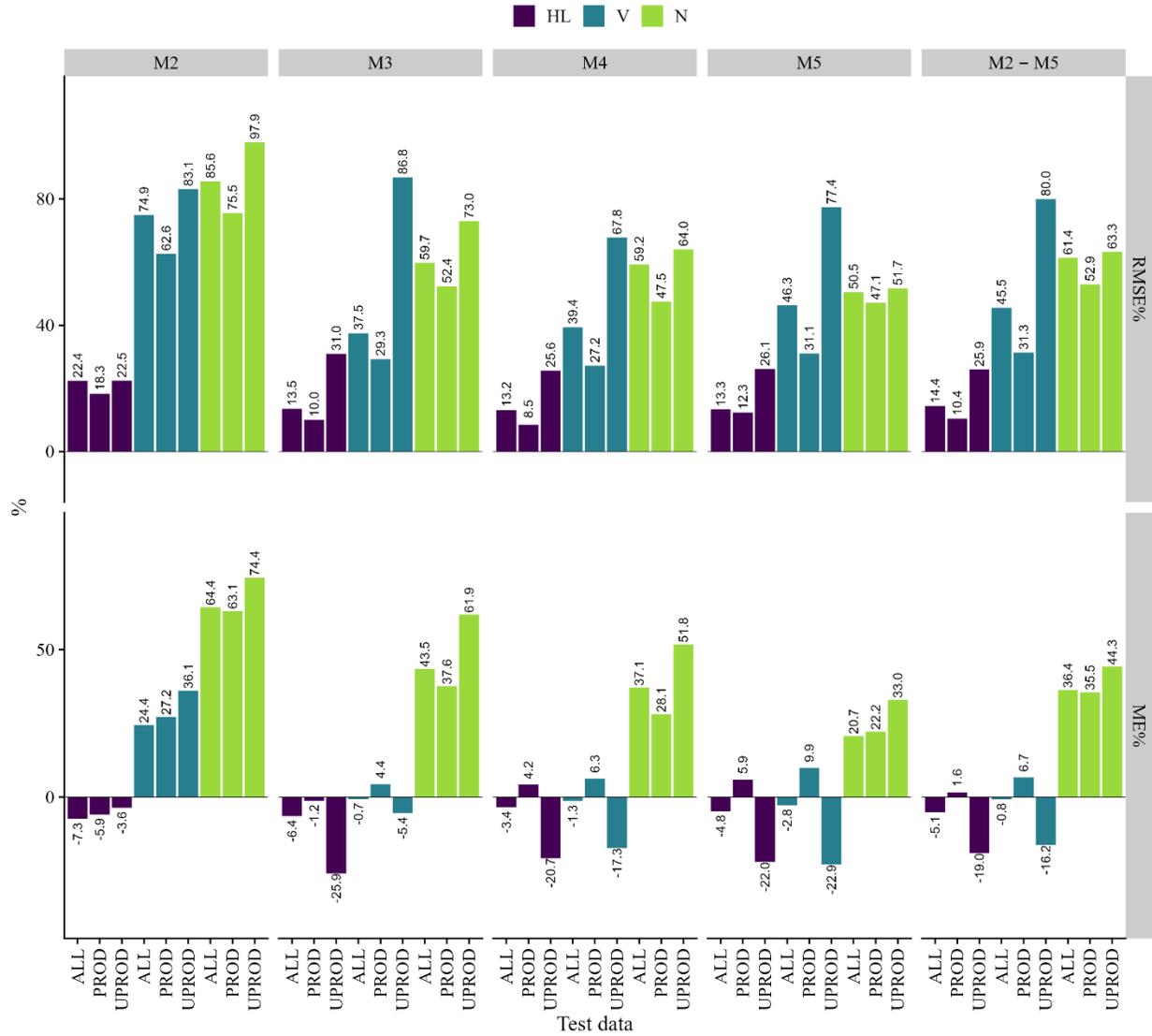

**Figure 6.** Relative root-mean-square errors (RMSE%) and mean errors (ME%) associated with the forest attribute predictions by maturity classes in the ALL, PROD and UPROD datasets. HL – Lorey's height, V – volume, N – stem frequency. Please refer to Section 2.2 for the maturity classification.

## 3.1 Large-area estimation of forest attributes

Characteristics associated with the large-area estimation of the forest attributes for forested land including all maturity classes (M1–M5) are presented in Table 5. Non-zero $\hat{\mu}_{cor}$ values indicate that the application of the harvester models for the large-area synthetic estimation of forest attributes would result in systematic errors if they are not corrected by means of a probability sample of field plots. The largest systematic errors were observed for N ($\hat{\mu}_{cor}$ = 39.0 % of the direct estimate). The smallest systematic error of the synthetic estimate was observed for V ($\hat{\mu}_{cor}$ = 0.8 % of the direct estimate).

The MA estimation using the harvester models resulted in efficiency gains, compared with the direct estimation, for all forest attributes (Table 5). The smallest RE value of 1.70 was associated with N whereas the largest RE value of 6.00 was associated with the estimation of HL. In this section, we presented results concerning forest attributes HL, V, and N, whereas the results for AGB, G, and QMD are in the Appendix.

**Table 5.** Characteristics associated with the estimation of forest attributes of the study area. V – volume, HL – Lorey's height, N – stem frequency, SE – standard error.

| Forest attribute | Direct estimate $\hat{\mu}$ | 2SE% $\hat{\mu}$ (%) | 2SE% $\hat{\mu}_{MA}$ (%) | Correction factor of the synthetic estimate $\hat{\mu}_{cor}$ (%) | Relative efficiency |
|---|---|---|---|---|---|
| HL (m) | 12.40 | 0.84 | 0.34 | -6.53 | 6.00 |
| V (m³ · ha⁻¹) | 124.09 | 2.16 | 1.14 | 0.82 | 3.58 |
| N (stems · ha⁻¹) | 957.44 | 1.58 | 1.37 | 38.98 | 1.70 |

## 4 Discussion

Harvesters operate in forests that are considered harvest-ready from the perspective of timber production. In Norway, forests that have reached an age of 60–120 years typically fall under this definition. The decision on final felling also depends on site fertility and tree species. It is also worth noting that forests managed for timber production may have characteristics that differ from those of unmanaged forests. For example, silvicultural operations that affect the size distribution of trees, such as thinnings from below, may be carried out in productive forests. The selection bias towards mature and productive forests associated with the training data of the harvester models manifests in larger systematic errors in both young forests and deciduous-dominated forests than in older coniferous forests. Deciduous species are currently rarely used for commercial timber production purposes in Norway, and deciduous trees were therefore underrepresented in our harvester dataset (see Fig. 2). We also observed that large-area synthetic estimates resulting from harvester-based models would have large systematic errors. However, systematic errors can also by chance balance out as was the case for timber volume in our study. Timber volume had a tendency to be underpredicted in productive forests (PROD) and overpredicted in unproductive forests (UPROD). In total over the whole study area, this resulted in a systematic error close to 1 %.

Non-harvested trees are also a possible source of systematic estimation error. The non-harvested trees may partly explain why N had particularly large |ME%| values. Harvester data typically do not include the smallest trees (DBH < 10 cm) that do not fulfill the dimension requirements of commercial timber assortments. In many cases, small trees are manually removed before the machine-based harvesting operation to increase the efficiency of the harvester. The timber volume of those small trees is negligible, but the harvester-based field observations of stem frequency and mean attributes, such as mean diameter, are likely to differ considerably compared to the use of traditional field measurements as the reference for ABA model fitting. Currently, harvesters are not capable of recording non-harvested trees, and therefore we were not able to observe the possible mismatches between the harvester and ALS data. Close-range remote sensing solutions, that are capable of measuring tree-level attributes rapidly and accurately in forest stands (Ahola et al. 2021), could open possibilities to also keep track on trees left out of harvest operations.

Studies using large harvester datasets as the field-reference for ALS-based forest inventories are rather rare. Since previous studies have mostly focused on the prediction of timber volume, we will constrain the discussion associated with the predictive performance of the harvester models to timber volume to enable the comparison of error levels. The collection of harvester data has, nonetheless, already begun to mature towards operational standards in Sweden (Söderberg et al. 2021). In the prediction of forest attributes, Söderberg et al. (2021) and Maltamo et al. (2019) reported smaller RMSEs (9–15 %) for merchantable timber volume compared with our results associated with timber volume, because their level of validation was the stand-level, while we validated our results at the plot level. The reduction of uncertainty is a known property when aggregating estimates from finer into coarser scales (Kotivuori et al. 2021). Saukkola et al. (2019) predicted timber volume at the stand-level and reported, depending on modeling unit sizes, RMSEs of 26–31 % that are larger than our RMSEs reported at the plot-level for older production forests. They used a limited set of harvester data in the training of the model that may have negatively affected the stand-level accuracy they reported. Hauglin et al. (2018) reported RMSEs of 19–22 % and 32–56 % for timber volume in high-productive and low-productive forests at the level of 400 m$^2$ grid cells, respectively. Those uncertainties are approximately in line with ours for timber volume in older managed forest (RMSEs of 27 and 31 %). Noordermeer et al. (2022) estimated timber volume at the harvested site (i.e., a cluster of adjacent stands) level with an RMSE% of 9–17 % depending on the modeling unit size and simulated tree positioning errors. For 400 m² grid cells, they reported cross-validated RMSEs of 18–22 % depending on simulated tree positioning errors, which are smaller than our RMSE in older production forest (RMSE of 27 %). A possible reason for this is that the data used by Noordermeer et al. (2022) came from fewer harvested sites and were less variable compared with our test data that comprised independent NFI plots.

Söderberg et al. (2021) reported systematic errors of 2–4 % for volume although the predictive models were validated in harvest-ready forests. Similarly, Hauglin et al. (2018) reported systematic prediction errors of 4–11 % for volume in a highly productive forest. We observed positive systematic errors in a similar range as Hauglin et al. (2018) for volume in older productive forests. However, systematic errors were negative for unproductive, and often deciduous-dominated forests. The close to zero systematic errors associated with all forests can thus largely be explained by the volume model's tendency to underpredict in spruce forests and overpredict in deciduous-dominated forests. The results of this study suggest that systematic errors associated with harvester model predictions are difficult to avoid even if forests to be mapped resemble harvest-ready stands.

The efficiency of an estimator can be evaluated using the RE (relative efficiency). In our case, the RE can be interpreted as the number of additional field sample plots that would be required to obtain the same precision in direct (field-data based) estimation as in MA estimation that takes advantage of the harvester model and ALS data. The RE of 1.70 for stem number thus suggests that 70 % more field data would be required to achieve the same precision without using harvester and ALS data. There exist few comparable studies that have carried out MA estimations using field and ALS data. Räty et al. (2021a) reported an RE value of 2.11 for stem frequency with the MA estimator that used a model constructed for the prediction of DBH distributions using Norwegian NFI and ALS data. Räty et al. (2021a) also reported a much smaller systematic error that was 1.4 % compared with 39.0 % in our study. We re-analyzed a dataset by Hauglin et al. (2021) used NFI and ALS data for large-scale mapping of forest attributes in Norway. We calculated REs of MA estimators of HL, V, AGB, and G (for details see Appendix A6). The data of Hauglin et al. (2021) resulted in considerably larger RE values (7.85 for HL, 6.21 for V, 5.72 for AGB, and 5.10 for G) compared with the MA estimates using harvester models and ALS data in our study (6.00 for HL, 3.58 for V, 3.27 for AGB, and 3.51 for G). These findings show that the combination of field plots and ALS data can be more advantageous than the combination of harvester and ALS data. However, the harvester data which were used here and in other studies so far are rather limited in terms of the number observations, areal coverage or positioning accuracy. Considerable improvements may therefore be expected in the future, if the availability of harvester data improves.

Inaccuracy associated with the geographic positions of trees limits the use of harvester data, and the positioning errors of trees are also a potential source of uncertainty in modeling and mapping (Noordermeer et al. 2022). In order to minimize the effect of positioning errors, we utilized modeling units that are up to four times larger than the area covered by the NFI plots. According to Saukkola et al. (2019), the increase in the size of modeling unit from 254 $m^2$ to 761 $m^2$ or larger moderates the uncertainty associated with geographic positions of trees. Noordermeer et al. (2022) also showed that prediction errors due to positioning errors of trees can be considerably reduced by increasing the size of modeling unit from 100 $m^2$ to 400 $m^2$. The positioning accuracy of harvested trees can be improved by modern positioning technology, and it has been shown that sub-meter accuracy can be achieved with a differential GNSS system coupled with harvester boom sensors (Hauglin et al. 2017; Noordermeer et al. 2021). To the best of our knowledge, sub-meter positioning systems are so far not used operationally. Therefore, current operational applications must cope with positioning errors, for example, by increasing the size of modeling units.

The resolution mismatch between harvested grid cells and NFI plots may account for prediction errors to some extent when the models are applied in the test data. Packalen et al. (2019) showed that the prediction errors tended to decrease when training plots were larger compared to test plots. They also reported that resolution mismatch caused systematic errors but concluded that the resolution dependency had a minor effect in most real-world use cases. We minimized the effect of resolution mismatch by preferring predictor variables that are resolution invariant, but it was not always reasonable in terms of predictive performance. For example, the predictive performance of the HL model would have been considerably degraded after omitting the non-resolution-invariant variable h95 from the model. Systematic errors in HL were nonetheless relatively small, confirming the observations by Packalen et al. (2019).

The time lag between the collection of ALS data and harvester data or NFI data was up to 9 years, which potentially caused some systematic errors when harvester-based models are applied in the NFI data. We used the $time_{diff}$ predictor variable to account for the time lags between the field measurements and ALS data acquisitions in the regression modeling (Hauglin et al. 2021). Although we ensured that the interpretation of the coefficient associated with $time_{diff}$ was reasonable, the approach is still suboptimal since it cannot account for local growing conditions, such as tree species or site index. Another possible source of systematic errors is the heterogenous ALS data. The ALS data were collected using several acquisitions with varying flight parameters. We pursued to minimize the effects of dissimilar acquisitions parameters by using ALS metrics computed based on first echoes that are not so sensitive on pulse penetration capabilities than last echoes.

This study shows that harvester data constitute a potential data source for the creation of forest attribute maps. However, systematic errors are likely to occur, and the magnitude of systematic errors may be challenging to ascertain based on harvester data alone. Systematic errors in the estimation of forest attributes for a region can be eliminated using MA estimation given that a comprehensive probability sample is available in the area of interest. Even maps with large systematic errors can then generate considerable efficiency gains compared to using traditional field data.

## Conclusions

We draw the following conclusions from this study:

1) The selection bias of harvesters towards harvest-ready sites negatively affects the predictive performance of harvester-based models in the prediction and mapping of forest attributes.

2) Synthetic (pixel-counting) estimates resulting from forest attribute maps created using harvester and airborne laser scanning data are likely to contain systematic errors. In some cases, systematic errors can, by chance, cancel out in large-area estimation, because harvester data-based models tend to behave differently in different forest types.

3) Systematic errors can be corrected using a model-assisted estimator if a comprehensive probability sample is available.

4) Model-assisted estimators, despite utilizing synthetic estimates with large systematic errors, can result in considerable efficiency gains compared with field data-based direct estimators. This can be used for reducing the field sampling intensity without increasing uncertainties or for decreasing uncertainties without increasing the field sampling intensity.

# Acknowledgements

This study was supported by the Norwegian research council through the PRECISION project – Precision forestry for improved resource use and reduced wood decay in Norwegian Forests (NFR 281140). We acknowledge Simon Berg for technical support with the harvester data.

**Data availability:** Data generated or analyzed during this study are not available due to the ownerships of the datasets.

**Competing interests:** The authors declare there are no competing interests.

# Appendix

**Appendix A1.** Observed versus predicted values associated with the harvester models in the training data. Harvester models refer to models that are fit using harvester information and airborne laser scanning data.

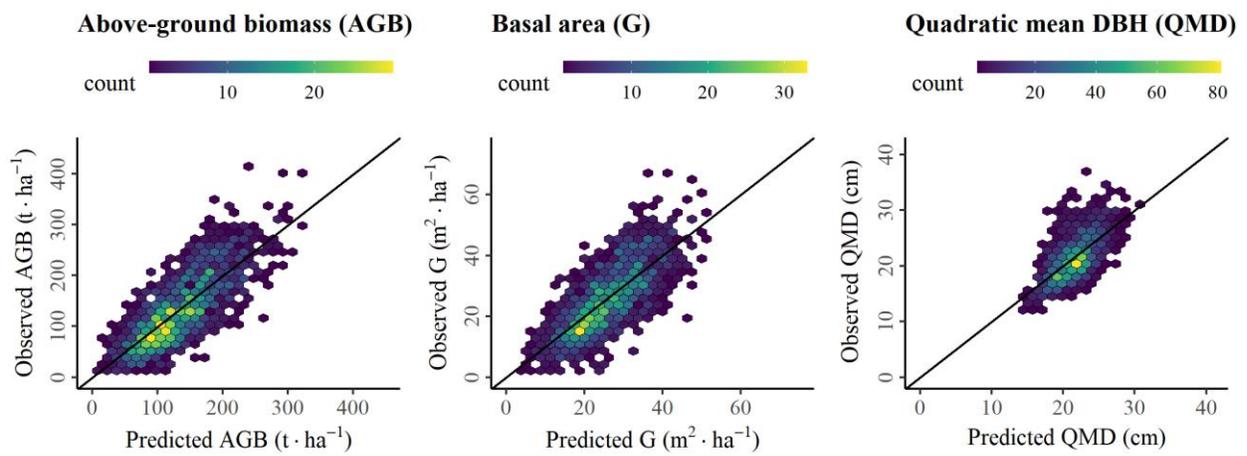

**Appendix A2.** Estimated coefficients, standard errors (SEs) of coefficients, p-values (p), relative root-mean-square errors (RMSE%) and coefficients of determination ($R^2$) associated with the forest attribute models fit using harvester and airborne laser scanning data.

|  | AGB | | | G | | | QMD | | |
|---|---|---|---|---|---|---|---|---|---|
| *Predictors* | *Estimates* | *SE* | *p* | *Estimates* | *SE* | *p* | *Estimates* | *SE* | *p* |
| (Intercept) | -45.41 | 5.44 | <0.001 | -6.22 | 0.97 | <0.001 | 10.99 | 0.70 | <0.001 |
| hmean | 15.97 | 0.66 | <0.001 | 1.82 | 0.12 | <0.001 | -0.79 | 0.09 | <0.001 |
| d2 | 57.16 | 12.46 | <0.001 | 23.78 | 2.23 | <0.001 | -2.60 | 1.05 | 0.013 |
| $time_{diff}$ | 5.24 | 0.44 | <0.001 | 0.66 | 0.08 | <0.001 | | | |
| h95 | | | | | | | 1.11 | 0.05 | <0.001 |
| RMSE% | 31.17 | | | 28.89 | | | 13.99 | | |
| $R^2$ | 0.62 | | | 0.56 | | | 0.41 | | |

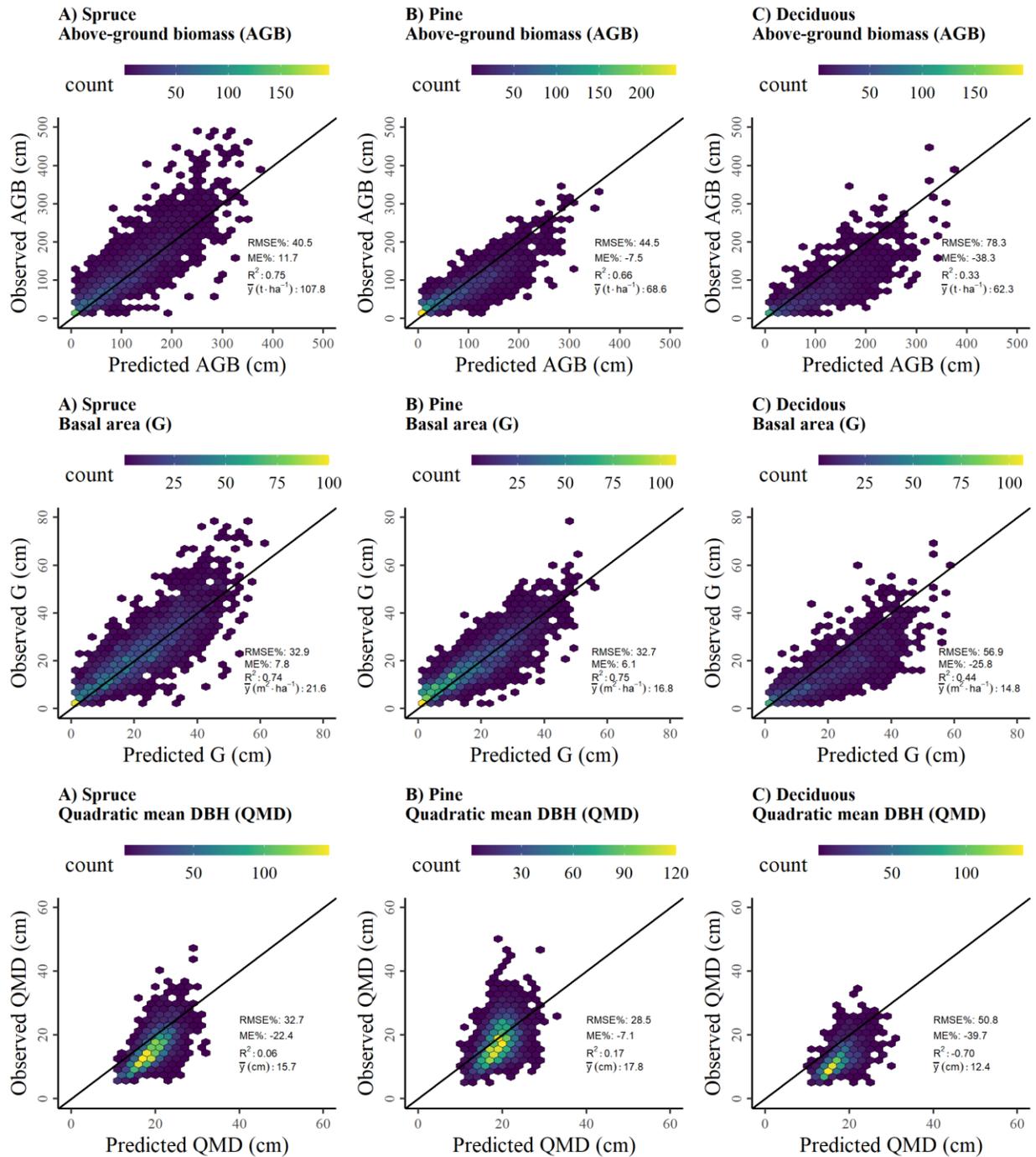

**Appendix A3**. Observed versus predicted values by dominant tree species (Spruce: Norway spruce, Pine: Scots pine, Deciduous: Deciduous species) in the ALL dataset.

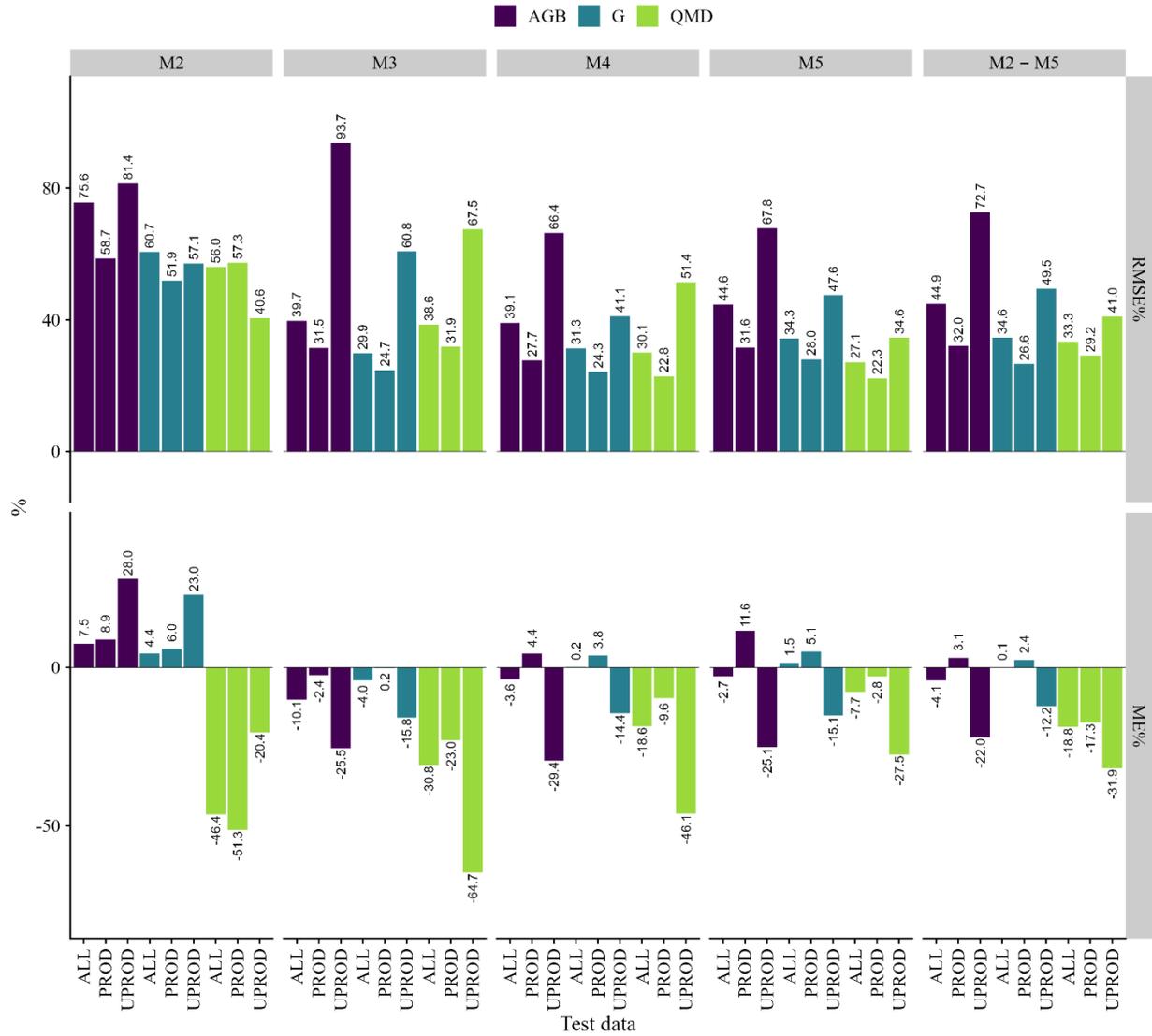

**Appendix A4.** Relative root-mean-square errors (RMSE%) and mean errors (ME%) associated with the forest attribute predictions by maturity classes in the ALL, PROD and UPROD test data. AGB – above-ground biomass, G – basal area, QMD – quadratic mean diameter.

**Appendix A5.** Characteristics associated with the estimation of forest attributes of the study area. AGB – above-ground biomass, G –basal area, QMD – quadratic mean diameter, MA – model-assisted, SE – standard error.

| Forest attribute | Direct estimate $\hat{\mu}$ | 2SE% $\hat{\mu}$ (%) | 2SE% $\hat{\mu}_{MA}$ (%) | Correction factor of the synthetic estimate $\hat{\mu}_{cor}$ (%) | Relative efficiency |
|---|---|---|---|---|---|
| AGB (t · ha⁻¹) | 80.5 | 1.98 | 1.08 | -2.47 | 3.27 |
| G (m2 · ha⁻¹) | 17.81 | 1.61 | 0.86 | 1.46 | 3.51 |
| QMD (cm) | 15.82 | 0.83 | 0.69 | -18.77 | 1.45 |

**Appendix A6.** Results of MA estimators using the data and models by Hauglin et al. (2021) who used Norwegian national forest inventory and airborne laser scanning data for forest attribute mapping. Note that the study by Hauglin et al. (2021) only covered a subset of the forest land area covered by our study. However, 86 % of the plots on the forest land are the same as in our study. The same ALS data were used in both studies. The calculations given in this table are based on 6743 sample plots. For descriptions of the estimators, see Section 2.7. MA – model-assisted, SE – standard error.

| Forest attribute | Direct estimate $\hat{\mu}$ | 2SE% $\hat{\mu}$ (%) | 2SE% $\hat{\mu}_{MA}$ (%) | Correction factor of the synthetic estimate $\hat{\mu}_{cor}$ (%) | Relative efficiency |
|---|---|---|---|---|---|
| HL (m) | 12.72 | 0.89 | 0.32 | 0.31 | 7.85 |
| V (m³ · ha⁻¹) | 136.46 | 2.28 | 0.91 | 0.57 | 6.21 |
| AGB (t · ha⁻¹) | 87.40 | 2.07 | 0.87 | 0.57 | 5.72 |
| G (m² · ha⁻¹) | 19.33 | 1.67 | 0.74 | 0.52 | 5.10 |